# Inert gas as electronic impurity in semiconductors: The case for active infrared absorption in silicon

Nian-Ke Chen[1,#], Yu-Chen Gao[1,#], Ji-Hong Zhao[1,*], Chun-Hao Li[1], Qi-Dai Chen[1], Hong-Bo Sun[2,*], Shengbai Zhang[3,*], and Xian-Bin Li[1,*]

[1]*State Key Laboratory of Integrated Optoelectronics, College of Electronic Science and Engineering, Jilin University, Changchun 130012, China*

[2]*State Key Lab of Precision Measurement Technology and Instruments, Department of Precision Instrument, Tsinghua University, Beijing 100084, China*

[3]*Department of Physics, Applied Physics, and Astronomy, Rensselaer Polytechnic Institute, Troy, New York 12180, USA*

Corresponding authors: lixianbin@jlu.edu.cn, or zhaojihong@jlu.edu.cn, hbsun@tsinghua.edu.cn, or zhangs9@rpi.edu

**Abstract**

Inert (noble gas) elements are extremely inactive to surrounding chemical environment and are frequently employed as protective gas in various semiconductor fabrication processes. In this work, we surprisingly discover that high doses of argon up to $10^{17}$-$10^{20}$ cm$^{-3}$ can be measured in silicon exposed by laser pulses even after 1300 days. First-principles calculations and molecular dynamics identify a unique argon-locking-vacancy (ALV) defect atomic model in silicon. The ALV defect is dynamically robust in contrast to the frequently moving pure Si vacancy. While argon is chemically inert, it readily modulates defect states of the occupied vacancy via steric repulsion and rattling motions, leading to significant band splitting within bandgap and thus strong infrared absorptions. Moreover, the repulsion between substitutional argon and dangling bonds results in shallow donors which explains the confusion of enhanced n-type carriers in experiments. The work paves a way of using noble gas element to produce active infrared absorption source for the non-heteroepitaxy photonic detectors directly on silicon wafer at infrared communication wavelength.

Silicon (Si) based optoelectronic devices is at the heart of optoelectronic industry owing to their ability for Si integration. Among them, photodetectors working at infrared (IR) communication wavelength ($\lambda$) of 1.31/1.55 μm are indispensable. However, due to the well-known problem of low absorption at $\lambda \geqslant$ 1.1 μm, corresponding to the Si bandgap, Si is powerless in communication applications. Often, a different semiconductor with a suitable bandgap is heterogeneously grown on Si. However, issues with heteroepitaxy such as lattice mismatch can reduce or even degrade the performance of the detectors [1]. Another way is to introduce IR absorption sources in Si. For example, gap states can be created inside Si by chalcogenide dopants with the help of ultrafast laser pulses to result in doped black silicon [2-6]. It has a strong IR absorption at $\lambda$ = 1.31/1.55 μm. However, such IR sources are often not stable enough for applications. For example, the IR absorption at 1.31/1.55 μm in black silicon can be significantly reduced by annealing at 775 K for half an hour [4].

In 2018, Zhao et al. reported a form of black Si, which was fabricated by nanosecond laser pulses without any intentional element dopant except for a protective Ar gas [7,8]. It was quite unexpected that the photodiode fabricated based on such a black Si has a high and stable photoresponsivity of 260 mA/W at 5 V at $\lambda$ = 1.31 μm [8], which paves the way for practical sensing by a Si detector at the IR communication wavelength. Argon is a noble gas widely used as a protective gas in the electronic industry. As a matter of fact, the name of argon is derived from a *Greek* word that means lazy or inactive. Due to its fully occupied valence band electronic shell with eight electrons, there is little chemical reaction between argon and other elements. As such, it is also expected that the Ar gas has no effect on the property of semiconductors.

In this work, we report the observation of very high concentration of Ar ($10^{17}$-$10^{20}$ cm$^{-3}$) in ultrafast laser-modified Si using the secondary-ion mass spectrometry (SIMS) measurement. First-principles calculations and molecular dynamic studies reveal the unique atomic and electronic properties of the Ar-doped Si to result in an unexpected and strong IR absorptions. While the pure Si vacancy can produce dangling-bond state within bandgap, it is movable and unstable. In contrast, Ar atom can lock the Si vacancy to form a dynamically stable defect complex even up to 900 K. Thanks to its full electronic shell, Ar protects the dangling electrons of vacancy and retains its gap states. Moreover, rattling motion and Coulomb repulsion of Ar atoms can lead to an enhanced structural distortion and the further splitting of defect energy levels within the bandgap. Unexpectedly, the repulsion between substitutional Ar and dangling electrons makes the defect a shallow donor, which explains the confusion of laser irradiation induced n-type doping effect. As a result, the inert Ar atom in fact acts as an electronic impurity and offers active and robust sub-bandgap IR absorption source for Si photodetector. This solves the long-term difficulty of high photoresponsivity of Si based detectors at IR communication wavelength. The physics behind shed new light on a general strategy of employing inert elements to raise performances of semiconductor devices.

The concentrations of Ar atoms are measured by dynamic secondary ion mass spectrometer (D-SIMS) in laser modified Si samples in argon protecting atmosphere. The fabrication of such samples were reported in our previous work [8]. The D-SIMS instrument is equipped with a Cameca IMS-4F device using 8 keV $Cs^+$ primary beam. Density-functional theory (DFT) calculations are performed using the VASP code [9,10], where the projector-augmented wave (PAW) pseudo potential and generalized-gradient approximation (GGA) exchange-correlation functional developed by Perdew, Burke and Ernzerh are adopted [11-13]. A Si supercell that contains 216 atoms are used to describe defect effects. The energy cutoff for plane-wave expansion is 380 eV. The 3×3×3 Monkhorst-Pack grids are used as Brillouin-zone sampling for static energy and property calculations, while the Γ point is used for structural relaxation and molecular dynamic (MD) simulations. The band structures of supercells are unfolded by the modified *VaspBandUnfolding* package [14]. Energy barriers are calculated using the climbing image nudged elastic band (c-NEB) method [15,16]. The structures and charge density are visualized by the VESTA code [17]. The positions of vacancy defect are determined by the Wigner-Seitz method in the OVITO code [18].

To figure out the actual role of Ar atoms in laser modified Si, we carefully analyze the dose of Ar by SIMS measurements in this work. Figure 1(a) displays Ar concentration for such a typical Si sample, which was fabricated previously by nanosecond laser in Ar atmosphere [8]. It is unexpected that even the sample has been made over 1300 days, a very high dose of Ar atoms can be detected as from ~ $4 \times 10^{21}$ cm$^{-3}$ at the surface to ~ $4 \times 10^{17}$ cm$^{-3}$ at 2 μm below the surface. We reevaluate the specific detectivity (D*) of the photodetector based on the Si sample [8] and compare it to those of non-silicon photodetectors at the IR wavelength of λ = 1.31 μm [19]. The 1.31-μm wavelength, whose photonic energy is below the bandgap of Si, is out of the detecting scope of detectors based on intrinsic Si. A higher D* reflects a higher signal-to-noise ratio of detectors. For example, Fig. 1(b) shows that the D* of the laser modified Si in Ar atmosphere here working at 295 K ($10^{11}$ cmHz$^{1/2}$W$^{-1}$) is not only higher than those of PbSe and InAs working at the same temperature but also higher than or close to those of InAs and InSb working at a much lower temperature of 193 K or 77 K. All of these indicate the inclusion of Ar in Si could potentially offer a highly effective IR absorption source for detectors.

To uncover the microscopic picture and critical role of Ar in Si samples, we carry out first-principles calculations. In fact, the formation energies ($\Delta H_f$) of Ar defects in crystalline Si are very large. Table S1 in Supplemental Material summarizes the calculated $\Delta H_f$ of several defects in Si, which agrees with previous reports [20,21]. The calculated $\Delta H_f$ of interstitial and substitutional Ar defects are as large as 6.05-7.19 eV while the $\Delta H_f$ of a silicon vacancy (V$_{Si}$) is about 3.67 eV. Despite of the high $\Delta H_f$, the Ar-related defects can still be formed under laser irradiations. A key reason is that the $\Delta H_f$ can be substantially reduced when Si is melted by laser irradiations, see Fig. 1(c). Also, compared with interstitial Ar defects, substitutional Ar defects should be dominate because an interstitial Ar and a V$_{Si}$ will be annihilated into a substitutional Ar once they encounter during the annealing process. This annihilation is obviously energy favorable. Moreover, the $\Delta H_f$ of substitutional Ar defects can further be lowered by their accumulation [see Fig. 1(d)] because such an accumulation reduces the number of dangling bonds.

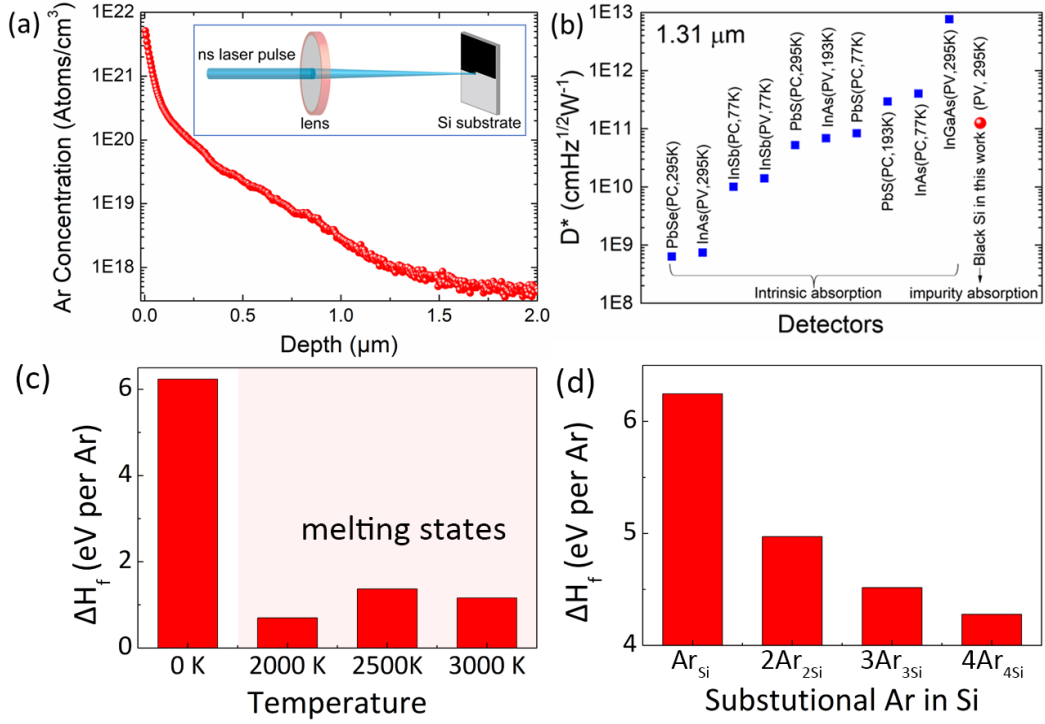

FIG. 1. (a) Concentration of Ar in the nanosecond-laser modified silicon sample measured by the secondary ion mass spectroscopy. (b) Comparison of specific detectivity ($D^*$) between the black silicon detector [8] and other reported infrared detectors at 1.31 μm [19]. (c) Formation energies of a substitutional Ar defect ($Ar_{Si}$) at various temperatures. The energy of a high-temperature state is calculated by the average free energy of the last 10 ps frames of a 20-ps *NVT* MD simulation. (d) Formation energies of the accumulated substitutional Ar defects in Si. The formation energies of the multi-substitutional defects are averaged by the number of Ar atoms.

Figure 2 elucidates atomic and electronic structures after an Ar atom is introduced into Si. In order to study Ar induced defect states in Si, a large supercell with 216 atoms is employed. Due to Brillouin Zone folding in the supercell, band structures are calculated and further reanalyzed by the effective band unfolding method [22]. The unfolded band structure in Fig. 2(d) reproduces the correct ***E-k*** dispersion for the intrinsic Si despite of underestimating the bandgap by the GGA-DFT method.

Naturally, an Ar atom filling in a Si lattice vacancy ($V_{Si}$) could be an ideal candidate for the IR absorption. Because an intrinsic vacancy has the ability of offering defect states within the bandgap. Fig. 2(a) displays the local tetrahedral motif in intrinsic Si. When a Si atom is removed to form a $V_{Si}$, four dangling bonds are produced, see Fig. 2(b). Here, the four atoms around the vacancy are noted as atoms 1, 2, 3, 4 ($Si_1$, $Si_2$, $Si_3$, $Si_4$), respectively. $L_{i,j}$ is the distance between any two of them. If no any atomic relaxation, the vacancy with all the same $L_{i,j}$ = 3.87 Å holds an ideal $T_d$ local symmetry. However, due to the well-known Jahn-Teller distortion effect, the four atoms are relaxed closer to the center of the vacancy with $L_{1,2}$ (3.03 Å) ≈ $L_{3,4}$ (3.12 Å) and other $L_{i,j}$ = 3.54 Å. The symmetry is thus lowered to a near $D_{2d}$ symmetry ($\sim D_{2d}$) shown in Fig. 2(b), which is consistent with a previous result [23]. Accordingly, the energy levels of defect states can split into two parts: two occupied levels in the lower part within the bandgap while two unoccupied levels in the upper part within the bandgap, see Fig. 2(e) of the band structure and its schematic drawing [Fig. 2(h)] for Si with a $\sim D_{2d}$ vacancy.

After Ar atoms are introduced into Si which is confirmed by our SIMS measurement, the local structures of $V_{Si}$ will be changed. For example, as shown in Fig. 2(c), the atomic distortion of $V_{Si}$ can be further enhanced to hold a $C_{2v}$ symmetry due to a Ar atom substitutionally filled in $V_{Si}$ ($Ar_{Si}$). In this case, the Ar atom is closer to $Si_3$ and $Si_4$, which makes $Si_3$ and $Si_4$ move away from each other. Meanwhile, $Si_1$ and $Si_2$ move closer to each other. Therefore, $L_{1,2}$ and $L_{3,4}$ are changed to 3.46 Å and 5.40 Å, respectively. The ~120° bond angles of $Si_3$ and $Si_4$ indicate the bonding type is changed to be $sp^2$-hybridization like while the $sp^3$-hybridization like bonding still remains for $Si_1$ and $Si_2$. Our calculations found that the filling of Ar into $V_{Si}$ costs about 2.56-eV extra energy. Such large an interaction energy and the large structural distortion suggest that Ar should have a significant repulsive interaction with surrounding dangling bonds of Si atoms, i.e., the steric repulsion effect. The steric-repulsive induced distortion (SRD) also leads to a larger bonding hierarchy and changes the defect states of $V_{Si}$. Accordingly, comparing to the case of $V_{Si}$, defect levels of $Ar_{Si}$ are further splitted, see Figs. 2(f) and (i): the energies of highest occupied defect orbital (HODO), lowest unoccupied defect orbital (LUDO) and LUDO+1 is raised and closer to conduction band while the energy of HODO-1 lowers into the valence band.

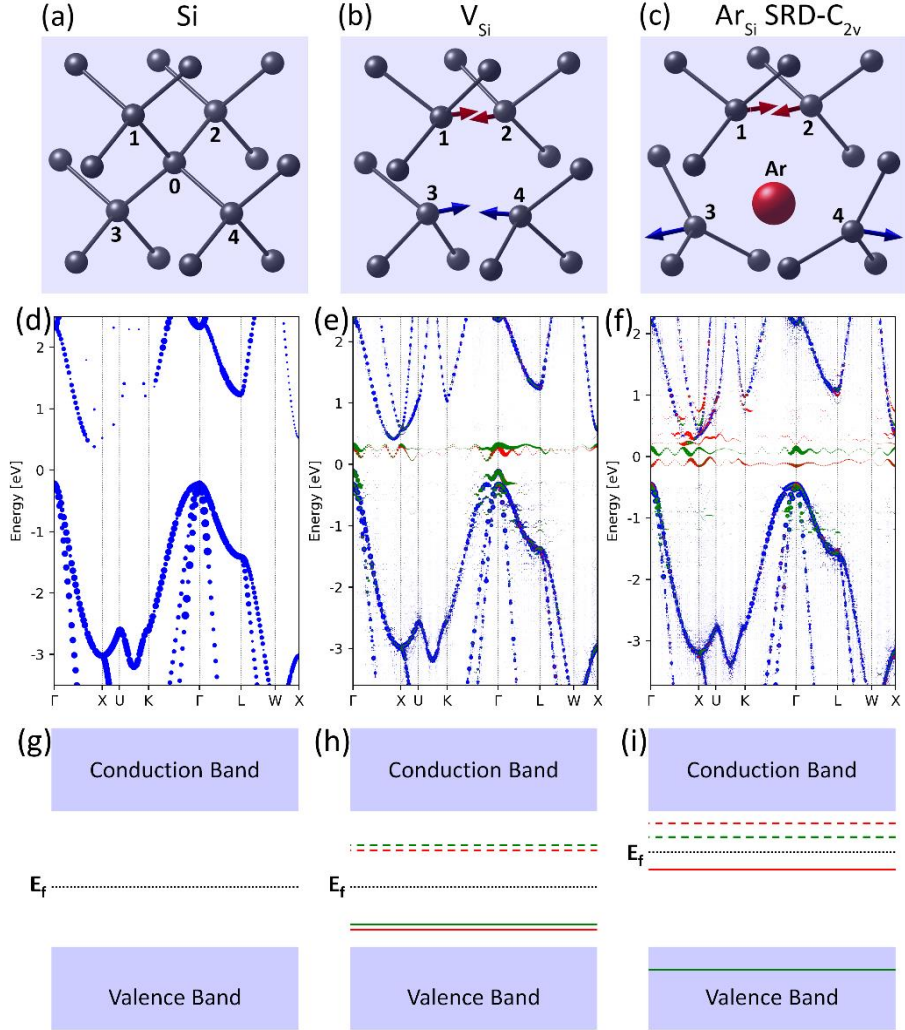

FIG. 2. Local atomic structures of (a) Si, (b) $V_{Si}$ and (c) $Ar_{Si}$ with the steric-repulsive distortion (SRD) holding a $C_{2v}$ symmetry. The arrows indicate the directions of atomic relaxation referred to their original positions. (d)-(f) correspondingly show the unfolded band structures of the three supercell models. The size of the scatters represents the atomic weight. The weights of $Si_{1,2,3,4}$ (green and red scatters) are amplified by a factor of 10. The green and red scatters indicate the contributions from $Si_{1,2}$ and $Si_{3,4}$, respectively. (g)-(i) show corresponding schematic band structures. $E_f$ is set as 0 eV.

In fact, the local atomic configuration of $Ar_{Si}$ defect with SRD is not unique. It can also have other structures with a $C_{3v}$ or $\sim T_d$ symmetry, see Figs. S1(e) and (f) in Supplemental Material. In both cases, the four dangling Si atoms ($Si_{1,2,3,4}$) are repulsed by the Ar and thus tend to have a sp$^2$-like bonding characteristic due to the bond angle of ~120°. In the two cases, not only the defect levels but also the spin states are splitted, see Figs. S2 and S3 in Supplemental Material. Three of the four electrons from dangling bonds occupy three spin-up levels while one occupies a spin-down level.

In fact, the three configurations of Ar$_{Si}$ with C$_{2v}$, C$_{3v}$ and ~T$_d$ SRDs almost have the same formation energy with little differnece of 0.01-0.03 eV (see Table S1 in Supplemental Material). The energy barrier of the tansition between C$_{2v}$ and C$_{3v}$ (C$_{3v}$ and ~T$_d$) configurations calculated by the NEB method is also as small as 0.015 (0.013) eV, indicating readily transitions to each other. Defect levels always preserve inside the bandgap after inert Ar doping in V$_{Si}$, despite of these different Ar configurations with close eneriges. Moreover, the SRD effect still stands in the accumulated substitutional Ar defects (3Ar$_{3Si}$ and 4Ar$_{4Si}$) and thus the splitted defect states also exist (see Fig. S4 in Supplemental Material). In addition, a bond-centered (B-C.) site interstitial Ar defect also has two dangling bonds while the tetrahedral/hexagonal-site interstitial Ar defects have no dangling bonds, see Figs. S1(a)-(c) in Supplemental Material. As such, the B-C. interstitial Ar also has two defect levels inside the bandgap, see Fig. S5 in Supplemental Material. All these defect states no doubt will benefit a robust sub-bandgap IR absorption in Si.

Next, electronic bonding mechanisms of Ar$_{Si}$ are analyzed to figure out the mechanism of the Ar doping induced change of defect states. Taking the C$_{2v}$ configuration as an example, the charge density difference (CDD) analyses [24] in Figs. 3(a)-(c) show no transferred or shared electrons between Ar and its surrounding Si due to the chemical inertia of Ar. Therefor, the interaction between Ar and the dangling Si atoms should be Coulomb repulsion effect which agrees with the SRD of the local structure. In fact, although Ar atom does not offer any defect levels but indirectly modulate defect states of a Si vacancy via the repulsion. Figures 3(d)-(h) elucidate the spatial distributions of defect states. The updated sp$^2$-like configuration of *Si$_{3,4}$* [see the bond angles of 121° in Fig. 3(a)] may release a dangling p$_z$ orbital from the orginal sp$^3$ orbital. Obviously, the LUDO+1 and the HODO are the unoccupied and occupied p$_z$-like orbitals of *Si$_{3,4}$*, respectively, see Figs. 3(e) and (g). Figure 3(f) shows the LUDO corresponding to the unoccupied sp$^3$-like orbital of *Si$_{1,2}$*. Since *Si$_1$* and *Si$_2$* has also a certain interaction due to a charge accumulation between them shown in Fig. 3(b), their occupied sp$^3$-like state further moves below the valence band maximum. We indeed find the state by projecting the orbital-decomposed partial charge density inside valence band [see Fig. 3(h)]. Such a defect state will move back into bandgap in the case of $C_{3v}$ and $\sim T_d$ SRDs due to the absence of such interaction between dangling Si atoms (see Figs. S2 and S3 in Supplemental Material).

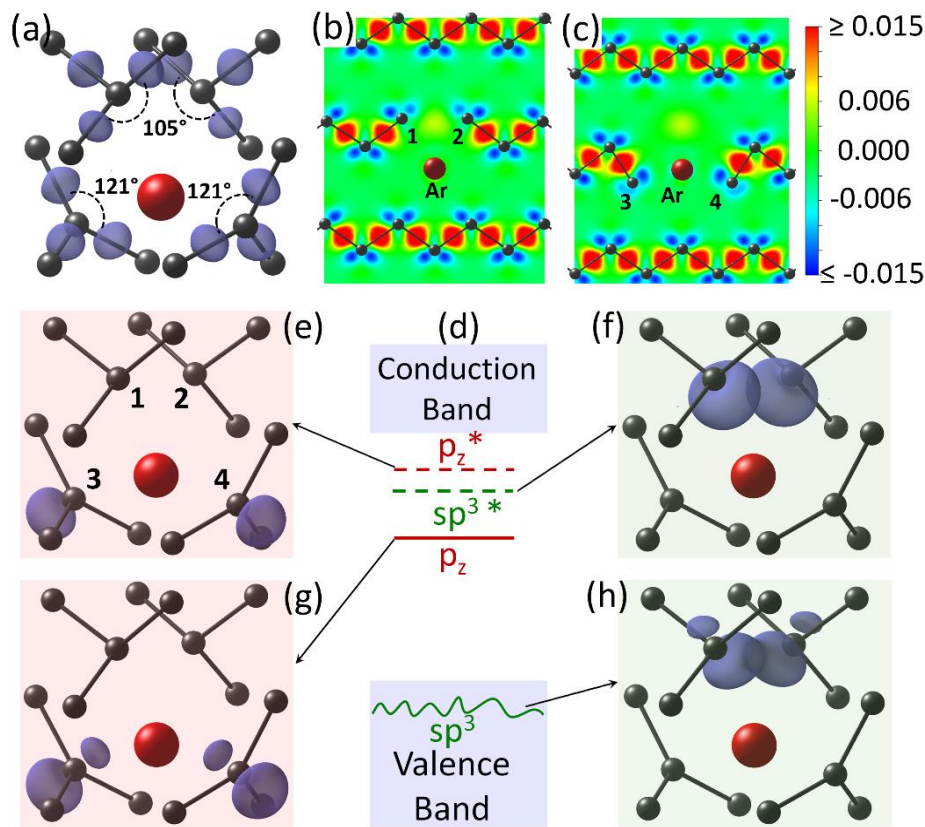

FIG. 3. (a) The charge density difference (CDD) of $Ar_{Si}$ with the $C_{2v}$ SRD. The value of isosurface is 0.015 $e/a_0^3$, $a_0$ is the Bohr radius. (b) and (c) show the respective $(1\bar{1}0)$ and $(110)$ slices of the CDD. The unit of the color bar is $e/a_0^3$. (d) Schematic defect levels of $Ar_{Si}$ with $C_{2v}$ SRD. (e)-(h) The corresponding orbital-decomposed partial charge density projected to real space. The values of isosurface are 0.005 $e/a_0^3$ for (e)-(g) and 0.0015 $e/a_0^3$ for (h).

To directly demonstrate the influence of the Ar-related defect on the IR absorption at $\lambda$ = 1.31/1.55 μm, the aobsorption coefficients are calculated [see Fig. 4(a)]. Here, the meta-GGA method using the modified Becke-Johnson (MBJ) exchange potential [25,26] is adopted to correct the bandgap underestimated by the traditional PBE functional (see Fig. S6 in Supplemental Material for the effect of correction). Obviously, a substantial enhancement of the sub-bandgap IR absorption is achieved by $Ar_{Si}$ in great contrast to the case without defect (Si) that displays almost no absoprtion. Moreover, the substitutioanl defects also have stronger abosorptions compared with the interstitial Ar and $V_{Si}$, owing to the repulsion induced splitting of defect levels.

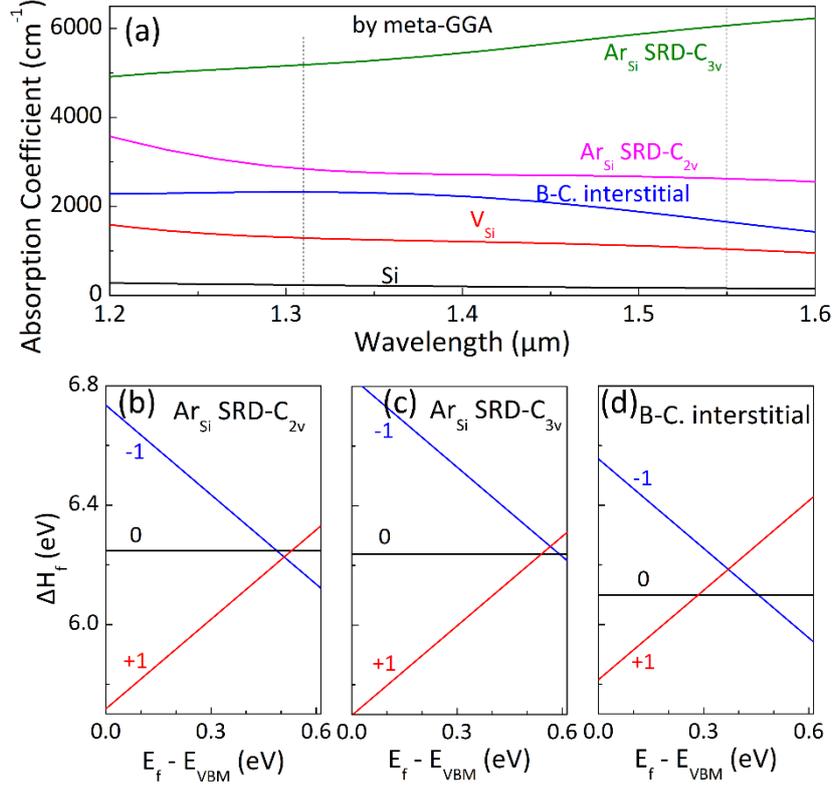

FIG. 4. (a) Calculated absorption coefficient of Si, $V_{Si}$, B-C. interstitial Ar and $Ar_{Si}$ with the $C_{2v}$ and $C_{3v}$ SRDs. (b)-(d) The $\Delta H_f$ of neutral and charged states of the $Ar_{Si}$ with $C_{2v}$ SRD, the $Ar_{Si}$ with $C_{3v}$ SRD and the B-C. interstitial Ar defect, respectively.

In fact, the repulsion effect of Ar on dangling electrons not only changes the atomic structure and the position of defect levels but also significantly affects the ionization process. When the Ar is filled into the vacancy, the strong repulsive interaction between Ar and dangling electrons should make the electrons be ionized easily. Figures 4(b)-(d) show the calculated chemical potential dependent $\Delta H_f$ of neutral and charged states of the Ar defects [27,28]. According to the position of transition levels (where the $\Delta H_f$ of neutral and charged states equal), it is surprising that the substitutional Ar defects are shallow donors but deep acceptors. In contrast, the B-C. interstitial Ar still has deep donor and acceptor levels. The shallow donor effect of $Ar_{Si}$ defects can explain the confusion that the fabricated sample shows a substantially enhanced n-type carrier concentration [8]. In other words, the substitutional Ar can act as an electronic impurity despite of its inert chemical property.

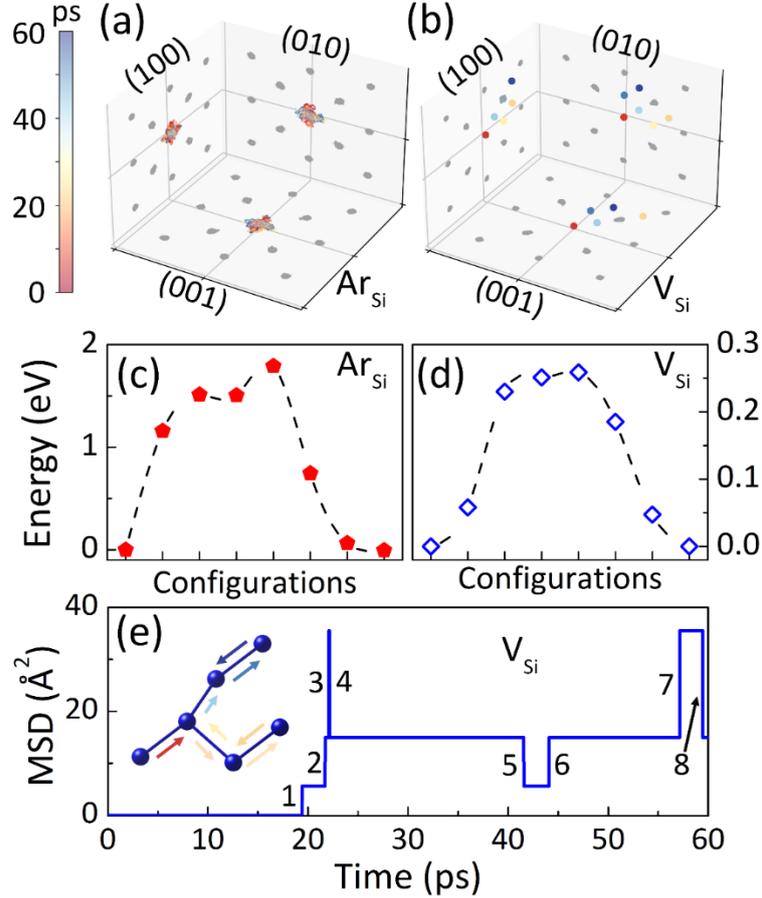

FIG. 5. The projected trajectories of the local structure of (a) $Ar_{Si}$ and (b) $V_{Si}$ during 60-ps MD simulations at 500 K. The colored dots represent positions of $Ar_{Si}$ and $V_{si}$ while the grey dots indicate the ones of Si atoms with time evolution. The energy barriers for a diffusion of (c) $Ar_{Si}$ and (d) $V_{Si}$ calculated by c-NEB method. (e) The MSD of the $V_{Si}$ during the MD simulation. Inset in (e) shows the schematic trajectory of diffusion of a pure $V_{Si}$ in the simulation.

It worth noting that $V_{Si}$ has the lowest $\Delta H_f$ (3.67 eV) among the defects in Si and it can also offer a significant sub-bandgap IR absorption. As mentioned above, the absorption of $Ar_{Si}$ is in fact originated from defect states of dangling bonds around the Ar filled vacancy. However, for practical device applications, IR absorption sources should be reliable enough. Therefore, two *ab initio* MD simulations of 60 ps at 500 K, which is significantly higher than room temperature at which detector devices work, are performed to compare the structural stability between $Ar_{Si}$ and pure $V_{Si}$. Figures 5(a) and (b) show the projected atomic trajectories with time for the local structures of $Ar_{Si}$ and the $V_{Si}$, respectively. During the MD simulations, the instant sites of the vacancy are determined by the Wigner-Seitz method as implemented in the OVITO code [18]. It is clear that the Ar atom in $Ar_{Si}$ is in form of a kind of rattling motion and moves much more intensely than normal Si atoms behave in lattices. The result is consistent with the negligible energy barriers among different SRD configurations of a $Ar_{Si}$. Despite of such intense motions, $Ar_{Si}$ stays in a same lattice firmly during the whole MD simulation. Note that the multi-substitutional Ar defects (e.g., $4Ar_{4Si}$) also stable during a 500 K MD simulation (see Fig. S7 in Supplemental Material). In a great contrast, the pure $V_{Si}$ shows an easy diffusion among different lattices. Next, the c-NEB analyses further evaluate barriers of atomic diffusions in Si. First, the energy barrier of a single $V_{Si}$ diffusion can be as low as 0.26 eV, see Fig. 5(d). The probability of its diffusion behavior at 500 K could be simply estimated by Arrhenius equation: $P = v \cdot \exp\left(-\frac{E_a}{k_B T}\right)$, where $k_B$ is the Boltzmann constant, $T$ is the temperature, $E_a$ is the energy barrier of diffusion and $v$ can be regarded as atomic vibration frequency. Using the frequency of highest optical phonon of crystalline Si, i.e., ~15 THz [29], the estimated characteristic time for one diffusion event is $\tau = \frac{1}{P} \approx 30$ ps for the $V_{Si}$. Figure 5(e) shows the mean square displacement (MSD) of $V_{Si}$ during the 500-K MD simulations. The pure $V_{Si}$ can move in different lattices as many as 8 times within 60 ps, which is close to the estimated characteristic time. That indicates in a practical Si sample the pure $V_{Si}$ can readily diffuses among the lattices at the raised temperature and will be eliminated when it gets to a grain boundary or a surface by an annealing process. As such, the significant sub-bandgap absorption by $V_{Si}$ as shown in Fig. 4(f) cannot readily happens. Second, in a great contrast, the energy barrier for an Ar diffusion out of a vacancy (i.e., an Ar exchanges its site with an adjacent Si atom) is as high as 1.8 eV, see Fig. 5(c). We have performed c-NEB calculations with other two different paths for Ar diffusions and the barriers are almost the same (~1.8 eV), see Fig. S8 in Supplemental Material. The atomic pictures of the Ar$_{Si}$ diffusions are presented in Figs. S9-S11 in Supplemental Material. According to the Arrhenius equation, the characteristic time for the Ar diffusion is at least ~ 70000 s, indicating a robust stability of Ar$_{Si}$. Moreover, in another 60-ps MD in Fig. S12 of Supplemental Material, the diffusion of Ar$_{Si}$ is still absent at a much higher temperature of 900 K, which is usually as an annealing temperature for fabrications of black Si detectors. Therefore, Ar$_{Si}$ can be regarded as a kind of Ar locking vacancy (ALV) defect.

Finally, we discuss benefits of the Ar$_{Si}$ or ALV doping for IR absorptions in Si detectors. Firstly, the steric repulsion induced band splitting makes the Ar$_{Si}$ defect readily contribute multi defect states and thus offer effective sub-bandgap IR sources, which is impossible for intrinsic Si. Secondly, due to Coulomb repulsion of the fully occupuied shell of Ar, local configurations of the ALV can be dynamically adjusted by the Ar atom at a raised temperature or even room temperature, see Fig. S13 in Supplemental Material, which leads to a broad IR absorption band. Thirdly, Ar acts like an electronic impurity with shallow donor defect levels. Then, the N$^+$ layer can be formed which is the key to construct the N+-N- junction in the device of IR detector [8].

**Conclusion**

In summary, we detect an unexpected high dose of inert Ar (with $10^{17}$-$10^{20}$ cm$^{-3}$) by SIMS measurements in laser modified Si samples at Ar protective gas even after more than 1300 days from when it was fabricated. First-principles calculations and molecular dynamics simulations demonstrate a mechanism of Ar locking vacancy (ALV or Ar$_{Si}$) happening in Ar doped silicon. While the pure vacancy in silicon can readily diffuse at 500 K, the ALV defect is dynamically robust at the same condition even no direct chemical bonding connection between Ar and its neighboring Si atoms. Despite of the chemical inert property of Ar, it can still act as an electronic impurity via strong Coulomb repulsion effect which leads to significant splitting of defect levels within bandgap, and thus have a strong sub-bandgap IR absorption in Si. It is an impossible task for intrinsic Si. Moreover, the repulsion between Ar and dangling electrons leads to a shallow donor effect, which explains the confusion of n-type doping effect of laser irradiation observed in previous experiments. It is reasonable to expect that such an inert element induced IR absorption mechanism may also happen in other semiconductors. Our work opens up a new door of using inert element doping engineering to develop high performance IR Si detector, which is urgently required in current Si based integrated optoelectronics.

## Acknowledgements

Work in China was supported by the National Natural Science Foundation of China (Grants No. 62275098, No. 12274180, No. 12274172) and the Fundamental Research Funds for the Central Universities. S.Z. was supported by the US Department of Energy under Award No. DE-SC0002623. We sincerely thank Prof. Q.Z. at USTC for his supports on band-unfolding analyses. The High-Performance Computing Center (HPCC) at Jilin University for computational resources is also acknowledged.

# Supplemental Material for
# "Inert gas as electronic impurity in semiconductors: The case for active infrared absorption in silicon"

Nian-Ke Chen[1,#], Yu-Chen Gao[1,#], Ji-Hong Zhao[1,*], Chun-Hao Li[1], Qi-Dai Chen[1], Hong-Bo Sun[2,*], Shengbai Zhang[3,*], and Xian-Bin Li[1,*]

[1]*State Key Laboratory of Integrated Optoelectronics, College of Electronic Science and Engineering, Jilin University, Changchun 130012, China*

[2]*State Key Lab of Precision Measurement Technology and Instruments, Department of Precision Instrument, Tsinghua University, Beijing 100084, China*

[3]*Department of Physics, Applied Physics, and Astronomy, Rensselaer Polytechnic Institute, Troy, New York 12180, USA*

Corresponding authors: lixianbin@jlu.edu.cn, or zhaojihong@jlu.edu.cn, hbsun@tsinghua.edu.cn, or zhangs9@rpi.edu

CONTENTS:

TABLE S1. Formation energies of the defects in crystalline Si. The unit is eV. $V_{Si}$ is a silicon vacancy. Tetrahedral (Tetra.), hexagonal (Hex.) and bond-centered (B-C.) represent three different sites of the interstitial Ar defects [see Fig. S1(a)-(c) for the atomic pictures]. $C_{2v}$, $C_{3v}$ and $\sim T_d$ represent three different configurations of the substitutional Ar defects as described in the main text [see Fig. S1(d)-(f) for the atomic pictures]. $2Ar_{2Si}$, $3Ar_{3Si}$ and $4Ar_{4Si}$ represent the accumulated multi-substitutional Ar defects [see Fig. S1(g)-(i) for the atomic pictures]. The formation energies of multi-substitutional Ar defects are averaged by the number of Ar atoms. Under the jellium approximation, the formation energy of defect $w$ with charge $q$ can be calculated by the following equation:

$$\Delta H_f(q,w) = E_{tot}(q,w) - E_{tot}(host) + \sum_i n_i \mu_i + q(\varepsilon_{VBM} + \varepsilon_F)$$

where $E_{tot}(q,w)$ is the total energy of the supercell with defects, $E_{tot}(host)$ is the energy without the defect (i.e., bulk Si), $n_i$ is the number of atoms being exchanged during the formation of the defect, $\mu_i$ is the atomic chemical potential of an element (here we use the $\mu_i$ of bulk and isolated atom for Si and Ar, respectively), and $\varepsilon_F$ is the Fermi energy relative to the valence band maximum, $\varepsilon_{VBM}$.

| $V_{Si}$ | Interstitial | | | Substitutional | | | Multi-substitutional | | |
|---|---|---|---|---|---|---|---|---|---|
| | Tetra. | Hex. | B-C. | $C_{2v}$ | $C_{3v}$ | $\sim T_d$ | $2Ar_{2Si}$ | $3Ar_{3Si}$ | $4Ar_{4Si}$ |
| 3.69 | 6.05 | 7.19 | 6.10 | 6.25 | 6.24 | 6.27 | 4.97 | 4.52 | 4.28 |

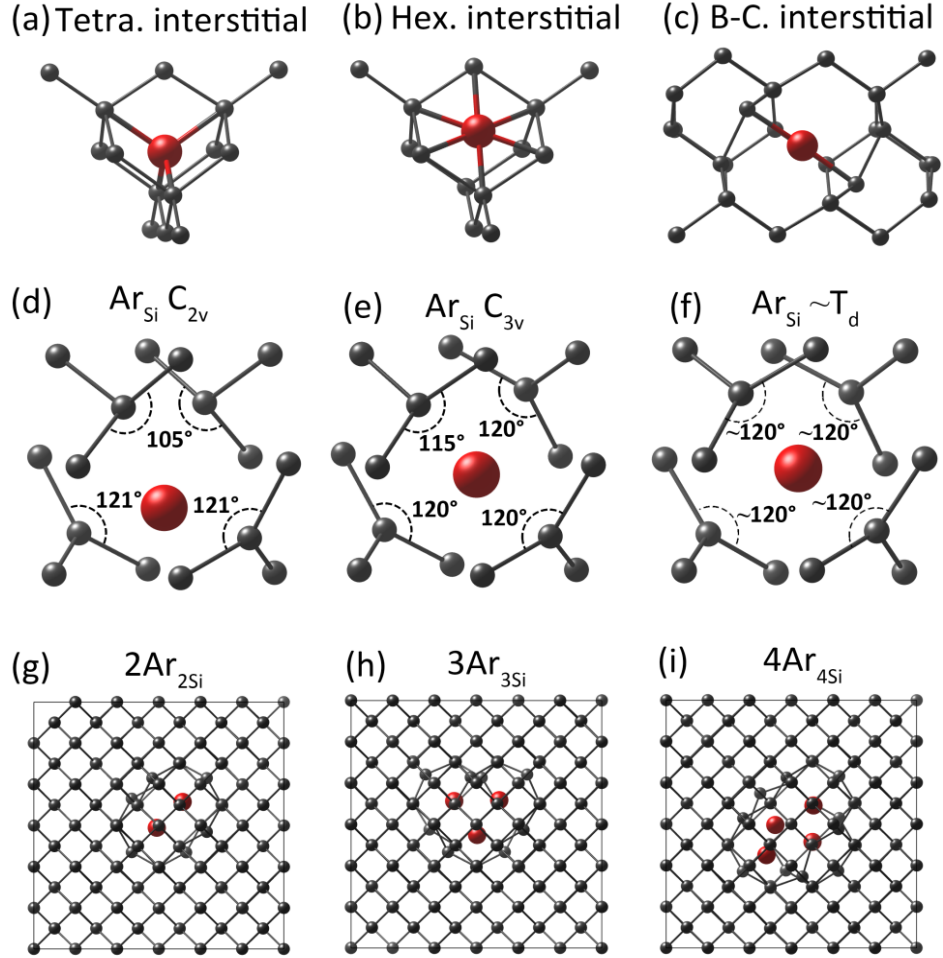

FIG. S1. (a)-(c) Local atomic structures of three interstitial Ar defects. (d)-(f) Local atomic structures of $Ar_{Si}$ with the steric-repulsive distortion (SRD) of $C_{2v}$, $C_{3v}$ and $\sim T_d$ symmetry. (g)-(i) The atomic structures of accumulated multi-Ar doping defects including two-Ar-atoms substitution ($2Ar_{2Si}$), three-Ar-atoms substitution ($3Ar_{3Si}$) and the four-Ar-atoms substitution ($4Ar_{4Si}$).

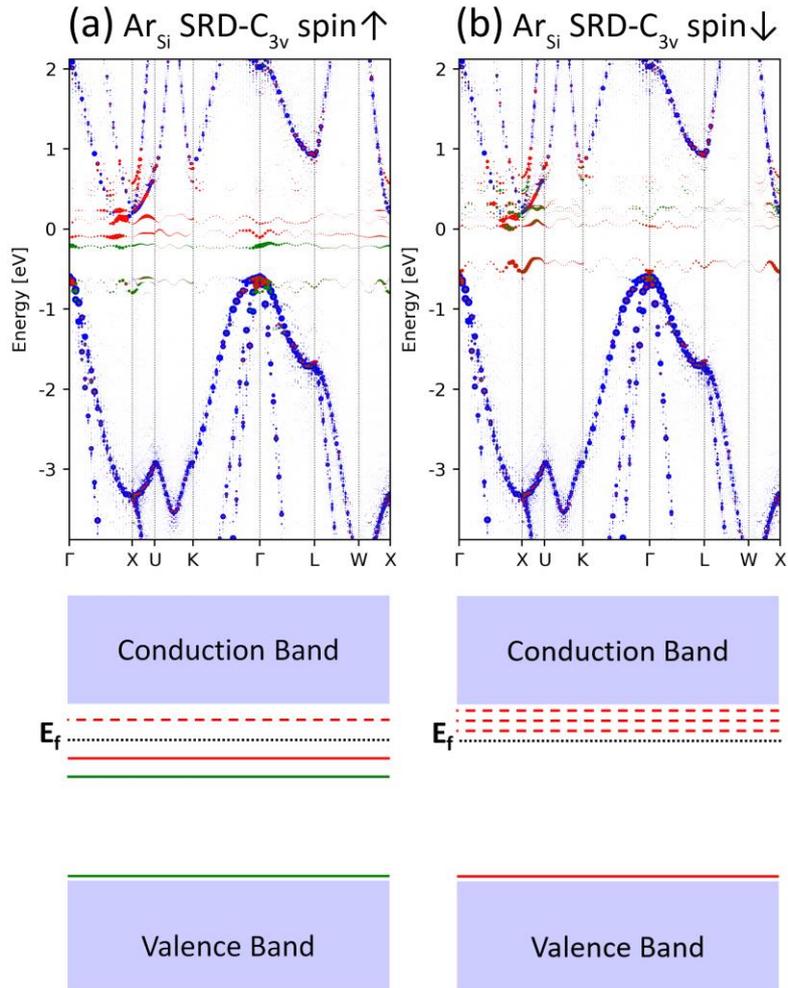

FIG. S2. Unfolded band structures of Ar$_{Si}$ with the SRD-C$_{3v}$ configuration [see Fig. S1(e) for the atomic structure] holding (a) spin-up and (b) spin-down polarizations. The green and red scatters indicate the contributions from Si$_1$ and Si$_{2,3,4}$, respectively. Lower panels show corresponding schematic band structures. E$_f$ is set as 0 eV.

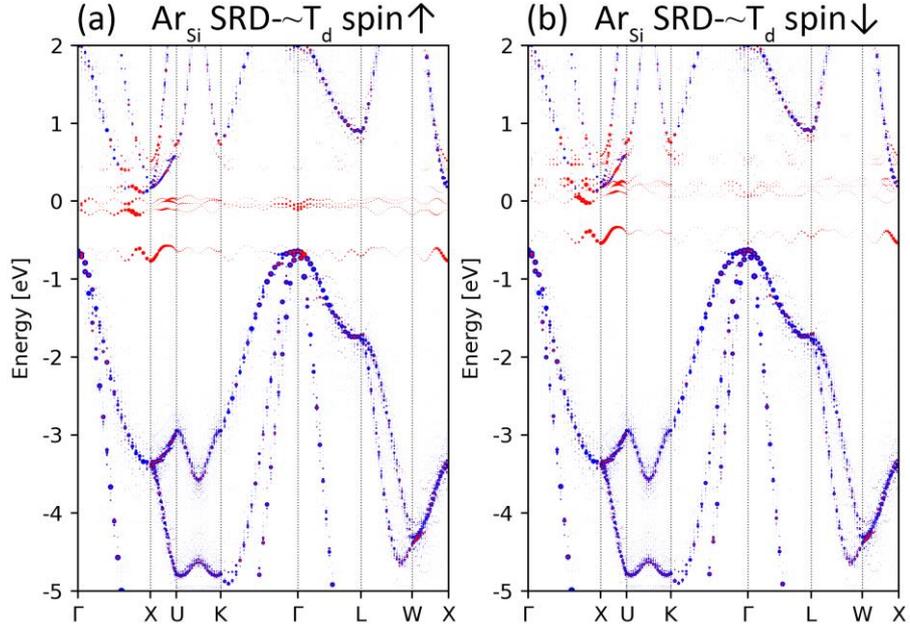

FIG. S3. The unfolded band structure of the Ar$_{Si}$ with the ~T$_d$ configuration [see Fig. S1(f) for the atomic structure], which is very similar to the Ar$_{Si}$ SRD-C$_{3v}$ configuration. The states contributed by the defective atoms are highlighted by red color.

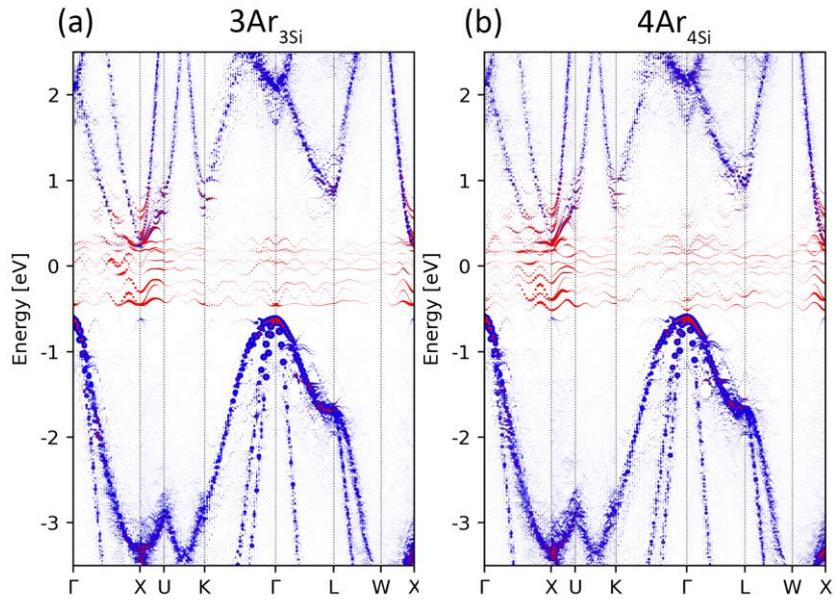

FIG. S4. The unfolded band structures of the (a) 3Ar$_{3Si}$ [see Fig. S1(h) for the atomic structure] and (b) 4Ar$_{4Si}$ [see Fig. S1(i) for the atomic structure], respectively. The states contributed by the defective atoms are highlighted by red color.

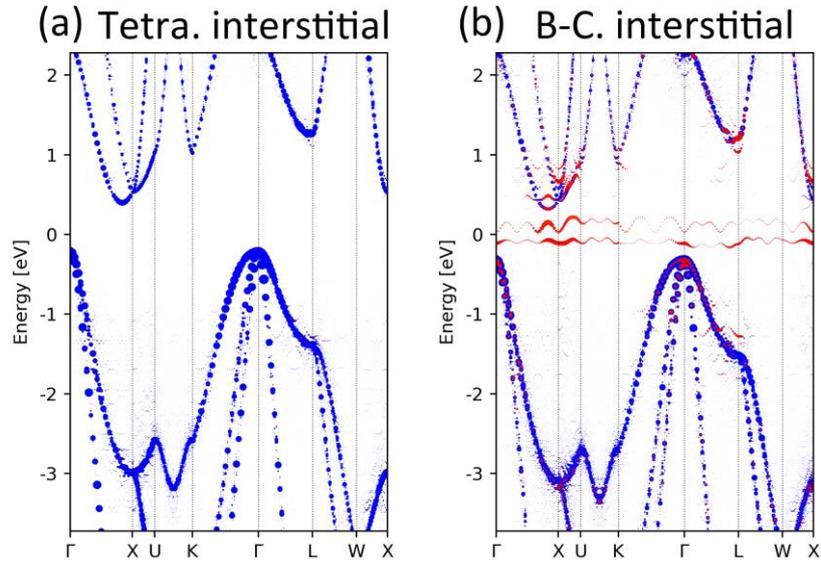

FIG. S5. The unfolded band structure of the (a) Tetra.-site [see Fig. S1(a) for the atomic picture] and (b) B-C.-site [see Fig. S1(c) for the atomic picture] interstitial Ar defects. The states contributed by the defective atoms are highlighted by red color. The defect states exist in the supercell with a B-C. interstitial Ar defect because the Ar atom in this case breaks a Si-Si bond.

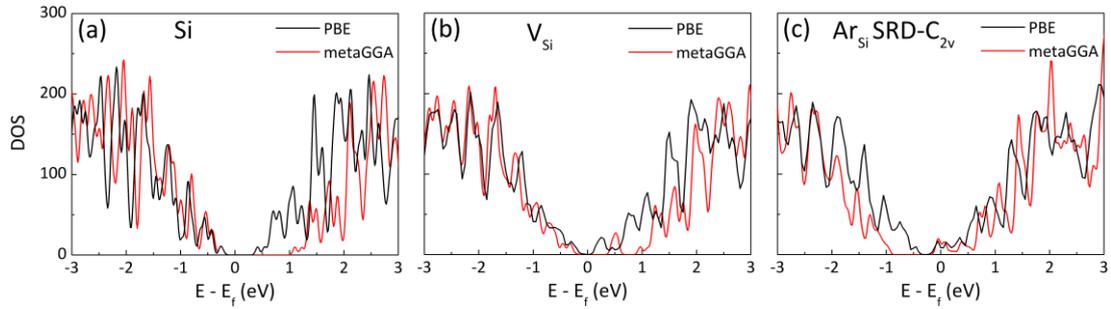

FIG. S6. Comparison of the density of states (DOS) of calculated by traditional PBE functional and the meta-GGA method using the modified Becke-Johnson (MBJ) exchange potential. The bandgap underestimated by PBE is corrected by the meta-GGA method while the defect states still retain.

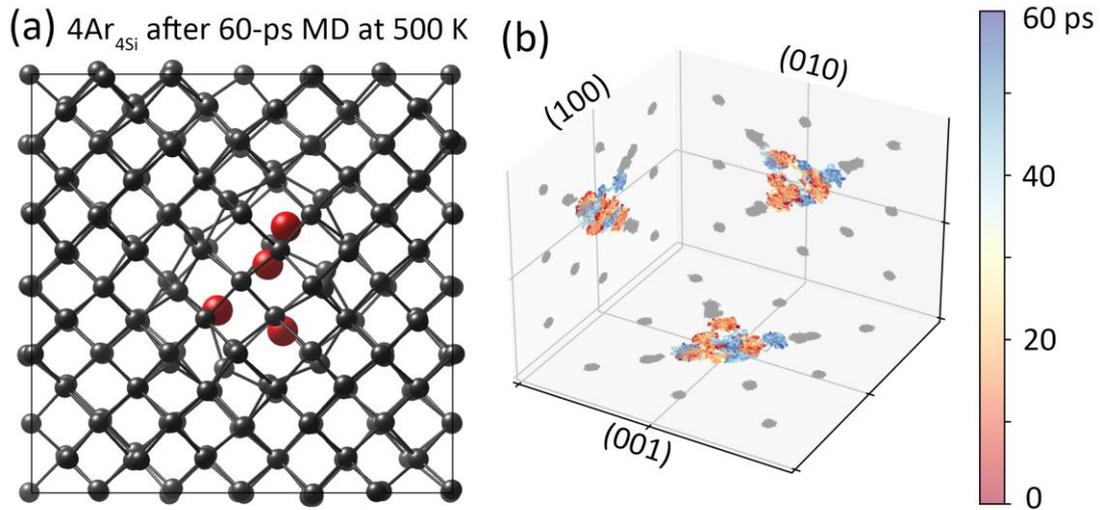

FIG. S7. (a) A snapshot of the atomic structure of the 4Ar$_{4Si}$ after the 60-ps MD simulations at 500 K. (b) The projected trajectories of the local structure of the 4Ar$_{4Si}$ during the 60-ps MD simulations at 500 K. The color coding is the same with that in the main text.

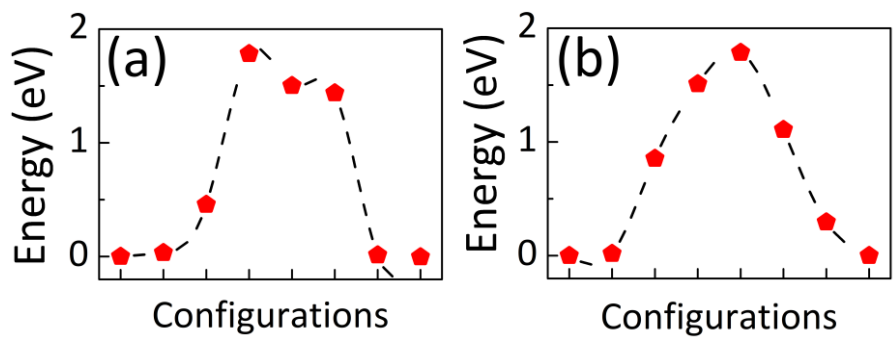

FIG. S8. The energy barriers for diffusion of Ar$_{Si}$ calculated by c-NEB method along different paths. The atomic pictures of different paths are shown in Figs. S9-S11 below.

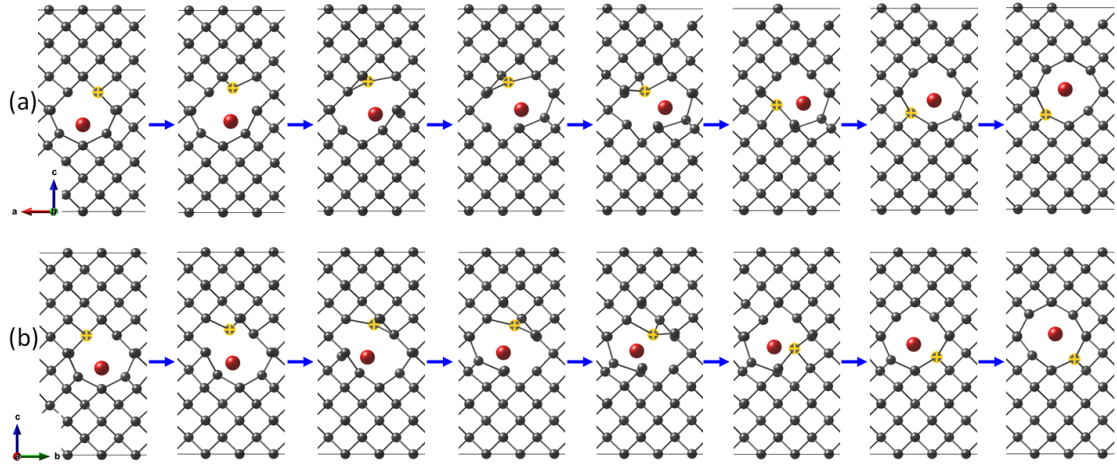

FIG. S9. The picture of Ar$_{Si}$ diffusion corresponding to the c-NEB calculation of the path in Fig. 5(c) of the main text. The color coding is the same with that in the main text. The yellow label indicates the Si atom who exchanged its position with Ar atom during the diffusion. (a) and (b) show the views along [010] and [100] directions, respectively.

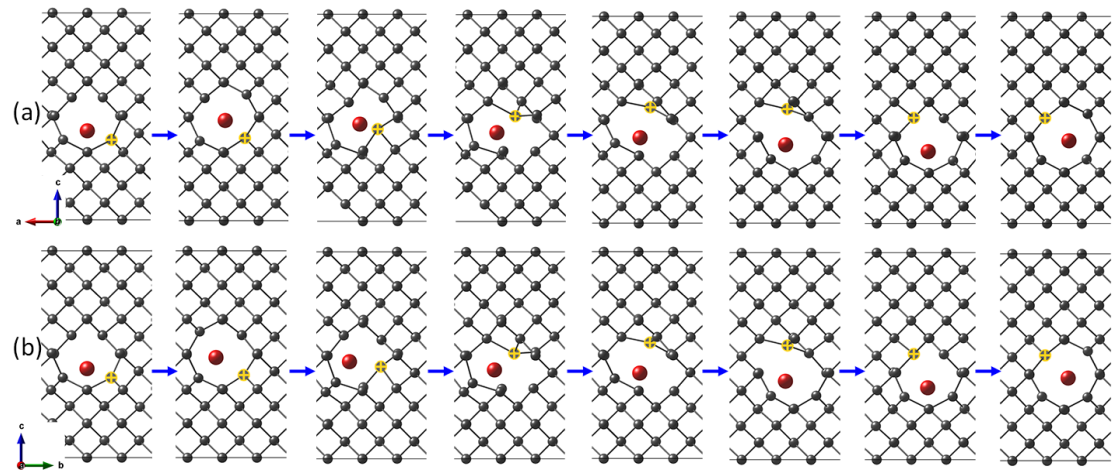

FIG. S10. The picture of Ar$_{Si}$ diffusion corresponding to the c-NEB calculation of the path in Fig. S8(a). The color coding is the same with that in the main text. The yellow label indicates the Si atom who exchanged its position with Ar atom during the diffusion. (a) and (b) show the views along [010] and [100] directions, respectively.

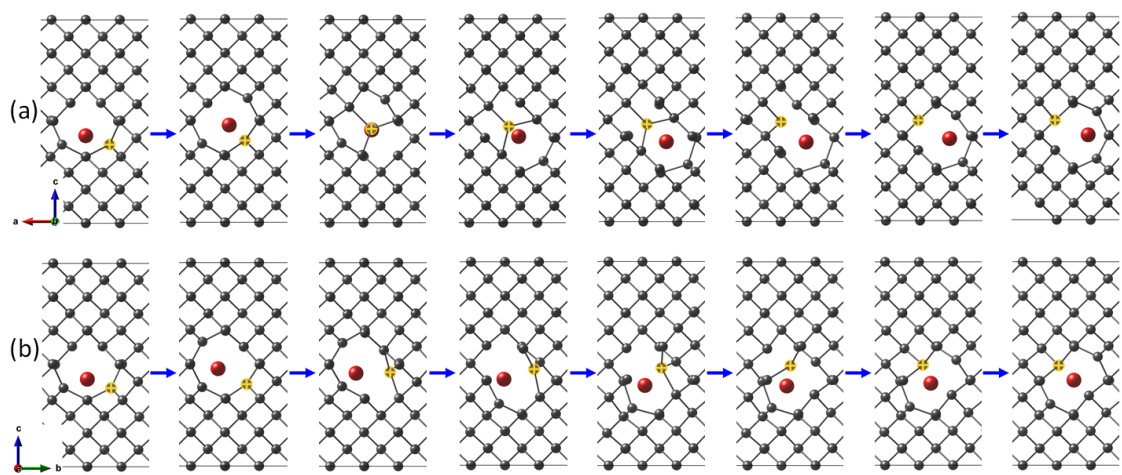

FIG. S11. The picture of $Ar_{Si}$ diffusion corresponding to the c-NEB calculation of the path in Fig. S8(b). The color coding is the same with that in the main text. The yellow label indicates the Si atom who exchanged its position with Ar atom during the diffusion. (a) and (b) show the views along [010] and [100] directions, respectively.

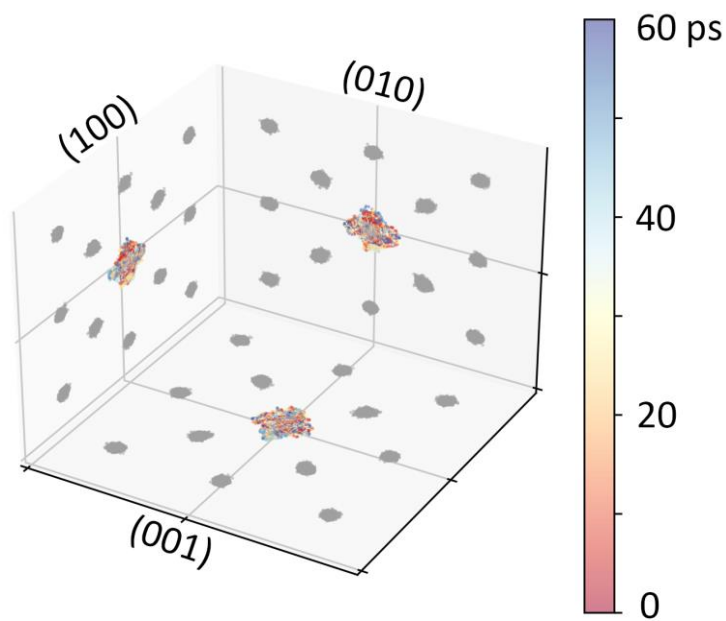

FIG. S12. The projected trajectories of the local structure of $Ar_{Si}$ during the 60-ps MD simulations at 900 K. The color coding is the same with that in the main text.

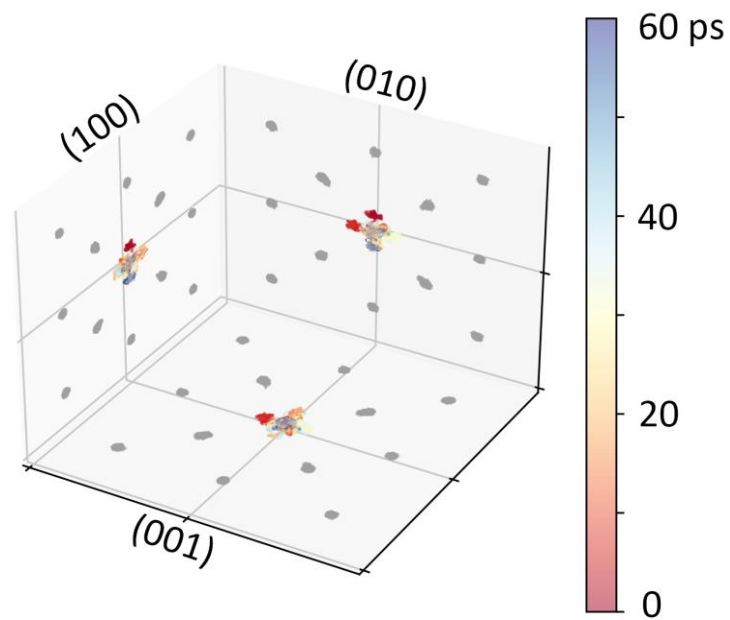

FIG. S13. The projected trajectories of the local structure of Ar$_{Si}$ during the 60-ps MD simulations at 300 K. The color coding is the same with that in the main text.